\documentclass{sig-alternate}
\usepackage{ams}

\def\sharedaffiliation{%
\end{tabular}
\begin{tabular}{c}}

\begin{document}
\title{Pre-Requirement Specification Traceability: Bridging the Complexity Gap through Capabilities}

\numberofauthors{3}
				\author{
			\alignauthor Ramya Ravichandar \\
     \email{ramyar@vt.edu}
       \alignauthor James D. Arthur \\
       \email{arthur@vt.edu}
		\alignauthor Manuel P\'erez-Qui\~nones \\
       \email{perez@cs.vt.edu}
       \sharedaffiliation
       \affaddr{Department of Computer Science}\\
       \affaddr{Virginia Polytechnic and State University}\\
       \affaddr{Blacksburg, Virginia 24060}
     }

\maketitle
\begin{abstract}
Pre-Requirement Specification traceability is the activity of capturing relations between requirements and their sources, in particular user needs. Requirements are formal technical specifications in the solution space; needs are natural language expressions codifying user expectations in the problem space. Current traceability techniques are challenged by the \textit{complexity gap} that results from the disparity between the spaces, and thereby, often neglect  traceability to and from requirements. 
We identify the existence of an intermediary region --- the transition space --- which structures the progression from needs to requirements. More specifically, our approach to developing change-tolerant systems, termed \textit{Capabilities Engineering}, identifies highly cohesive, minimally coupled, optimized functional abstractions called \textit{Capabilities} in the transition space. These Capabilities link the problem and solution spaces through directives (entities derived from user needs). Directives connect the problem and transition spaces; Capabilities link the transition and solution spaces. Furthermore, the \textit{process} of Capabilities Engineering addresses specific traceability challenges. It supports the evolution of traces, provides semantic and structural information about dependencies, incorporates human factors, generates traceability relations with negligible overhead, and thereby, fosters pre-Requirement Specification traceability. 
\end{abstract}

\category{D.2.1}{Software Engineering}{Requirements/Specifications}[]

\terms{Theory}
\keywords{Capabilities Engineering, Capabilities, Traceability, Requirements} 

\section{Introduction}
\label{sec:Introduction}

Current traceability techniques falter when subjected to the dynamics of requirements evolution, user needs volatility, market vagaries, technology advancements and other change-inducing factors that plague software systems during their development cycles. Undoubtedly, the inability to precisely capture and represent the effect(s) of each change has contributed to the long history of system failures 
\cite{Glass1998}. Although, the importance of traceability, and in particular requirements traceability (RT), was recognized in the early nineties \cite{Gotel1994}, the research community is still grappling with the challenges of traceability, especially between requirements and their source needs. In large part, this is because of the difficulty in formalizing user needs, which are often unstructured information. Although, needs are informal expectations of users, they serve as the primary source for requirements specification. If a software system is to exhibit the ``right'' functionality, then it is imperative that each system requirement satisfies one or more user needs.

Because system validation is performed against requirements, it is critical that we have the ability to trace requirements back to their source needs.
This type of traceability is known as pre-Requirement Specification (pre-RS) tracing \cite{Gotel1994}. In fact, it is empirically established that inadequate pre-RS tracing is far more responsible than than post-RS tracing (tracing requirements to design/code artifacts) for defective RT \cite{Finkelstein1991} \cite{Orlena}. We conjecture that this is because post-RS tracing works within the convenience of the \textit{solution space}, mapping requirements to design or code entities. 

\begin{figure}[htbp]
\centering
% trim = left bottom right top
\includegraphics[trim = 30mm 160mm 0mm 20mm, clip, width=10 cm]{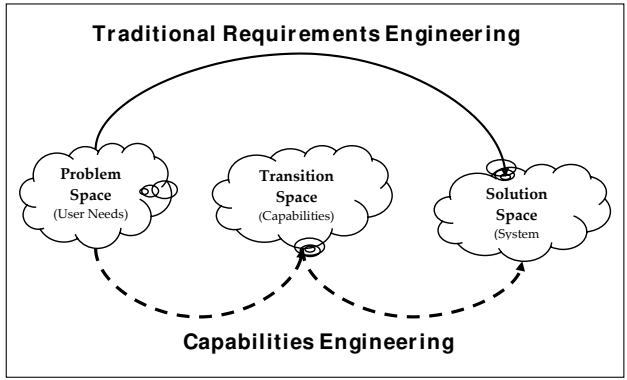}
\caption{\em Transition from Needs to Requirements}
\label{fig:Fig-ComplexityGap}
\end{figure}

However, pre-RS tracing is required to trace entities from the \textit{problem space} (needs) to the solution space (requirements). The chasm between these spaces, \textit{i.e.} the complexity gap \cite{Racoon1995}, is too large of a leap for current traceability techniques. In addition, the effects of this gap --- manifested as loss of domain information, misinterpreted requirements, misconstrued needs --- are exacerbated during the development of large-scale systems. We term the intermediary region that includes the complexity gap  as the \textit{transition space} and use it to ease the giant leap from needs to requirements. Figure \ref{fig:Fig-ComplexityGap} illustrates the difference between the traditional requirements engineering approach and our solution approach in advancing from the problem to the solution space.
%-------------------------------------------------------------------------- 		
%	FIGURE: Complexity Gap
%--------------------------------------------------------------------------
%---------------------------------------------------------------------------- 
%END FIGURE: Complexity Gap
%-----------------------------------------------------------------------------

Our solution approach for pre-RS traceability is derived from \textit{Capabilities Engineering} (CE), a process for architecting change-tolerant systems by constructing functional abstractions termed \textit{Capabilities} in the transition space \cite{Ravichandar2007}. These abstractions exhibit high cohesion and low coupling. The characteristic of high cohesion localizes the impact of a change. Low coupling implies reduced dependencies, and thereby, minimizes the ripple effect of change. Ripple effect is the phenomenon of propagation of change from the affected source to its dependent constituents \cite{Haney1972}. Capabilities influence the basic composition of systems, and in some sense, impose a high-level architecture. Capabilities are formulated from user needs and mapped to system requirements. In the process, they occupy a position that is neither in the problem space nor in the solution space. More specifically, although Capabilities are derived from user needs, they are imbued with design characteristics of cohesion and coupling. This introduces aspects of a solution formulation, and thus, discourages the membership of a Capability in the problem space. On the other hand, Capabilities are less detailed than entities that belong to the solution space. Consequently, Capabilities fit more naturally in the transition space. Furthermore, their formulation from the user needs and mapping to requirements imply that they have the potential to bridge the complexity gap; thus assisting the traceability between needs and requirements as depicted in Figure \ref{fig:Fig-ComplexityGap}. Moreover, the inherent ability of  Capabilities-based systems to accommodate change with minimum impact enhances the efficacy of traceability; random, unstructured ripple-effect impairs the strength of regular traceability techniques.

The Capabilities-based development approach strives to accommodate change with minimum impact, and therefore, incorporates pre-RS tracing in its process; traceability is the cornerstone of change-management. The use of the transition space facilitates the capture of domain information, and preserves relationships among needs and their associated functionalities during the progression between spaces. On the other hand, the characteristics of high cohesion and low coupling of Capabilities, support traceability in evolving systems by localizing and minimizing the impact of change. The ability to trace is unhindered by the system magnitude when utilizing a Capabilities-based development approach because traceability techniques are \textit{embedded into the process}. Moreover, by considering traceability as an integral part of the development effort, several issues relating to human factors can be resolved. For example, often the importance of traceability is undermined because it is regarded as a time-consuming activity when executed in isolation. 

The remainder of the paper is organized as follows: Section \ref{sec:Background} discusses related work and outlines the overall process of engineering Capabilities. In Section \ref{sec:CapabilitiesEngineering}, we define each space, discuss the role of CE in facilitating traceability within each space and examine the spaces' connectivity in terms of common linkages. Then, in Section \ref{sec:AddressingSpecificGCTChallenges} we present how  the use of a Capabilities-based development approach addresses specific challenges of traceability. Our conclusions are surmised in Section \ref{sec:Conclusion}.

\section{Background}
\label{sec:Background}
We first clarify the usage of the terms ``requirement'' and ``requirement specification'' in the context of our research. A common definition of RT is the ``ability to describe and follow the life of a requirement, in both a forwards and backwards direction'' \cite{Gotel1994}. Forward traceability is tracing a requirement to its design or code entities, and backward traceability is tracing a requirement to its sources \cite{Wieringa1995}. According to these definitions a requirement is not necessarily a part of a specification document. Pre-RS tracing, however, refers to the traceability aspects of a requirement before its inclusion in a specification document. Consequently, there is an overlap between the different modes of traceability; this is graphically illustrated in \cite{Pinheiro2003}. This overlap stems from the fact that a requirement is iteratively refined until it is suitable for a formal specification. As a result, any statement that describes what is expected from the system, irrespective of its level of refinement, is termed a requirement. However, we make a distinction with respect to the terminology used. We consider a requirement as a statement that is formally recorded in a software requirements specification document \cite{Parnas1986}. Therefore, we consider the terms requirement and requirement specification  as one and the same, and so use them interchangeably.

Several models have been constructed to assist pre-RS traceability. For example, Contribution Structures \cite{Gotel1995} consider the role of users, personnel, and others, in eliciting information to trace the origin of requirements. Similarly, Yu and Mylopoulos \cite{Yu1994} factor in the influence of the relationships between stakeholders on the RT process. These methods emphasize the social aspect of requirements elicitation.  On the other hand, Pohl \cite{Pohl1996} tailors the Requirements Engineering (RE) environment to capture traceability information between needs and requirements using PRO-ART. More general reference models for traceability have been developed by Ramesh and Jarke \cite{Ramesh2001}.
In the Capabilities-based development approach, traceability information is more of an implicit by-product. 

Empirical research evidence indicates the failure of traditional RE to cope with requirements evolution, especially when building large systems \cite{Bell1976}. In contrast to traditional RE which attempts to \textit{minimize} change, CE strives to \textit{accommodate} change with minimum impact. The difference in the change-management strategies explains why the activity of traceability is an inherent part of the CE approach, whereas it is considered an extraneous activity in the RE approach. Although, CE utilizes stable Capabilities to develop the system, it is imperative that the process also has the ability to trace to and from the changed entities inorder to ensure the successful accommodation of change. 

%-------------------------------------------------------------------------- 		
%	FIGURE: Capabilities Engineering Process
%--------------------------------------------------------------------------
\begin{figure}[htbp]
\centering
% trim = left bottom right top
\includegraphics[trim = 8mm 205mm 0mm 13mm, clip, width=10 cm]{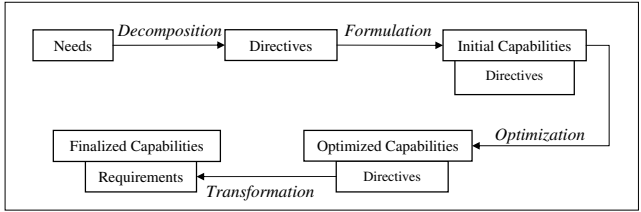}
\caption{\em Capabilities Engineering Process}
\label{fig:FigNeedsToRequirements}
\end{figure}
%---------------------------------------------------------------------------- 
%END FIGURE: Capabilities Engineering Phase
%-----------------------------------------------------------------------------
Figure \ref{fig:FigNeedsToRequirements} illustrates the major steps of the CE process. Capabilities are identified after needs analysis but prior to requirements specification, and indicate the functionality desired of a system at various levels of abstraction. As shown in Figure \ref{fig:FigNeedsToRequirements}, Capabilities are formulated from \textit{directives}. Directives are system characteristics obtained from user needs. They assist in the transfer of domain information, and help determine the initial sets of Capabilities, based on the values of cohesion, coupling, and abstraction level \cite{Ravichandar2007a}. These sets are then subjected to the constraints of technology and schedule to produce an optimized set of Capabilities. Throughout the process, Capabilities are associated with a set of directives, which are finally transformed to requirements. The output of the CE process is a set of finalized Capabilities, and their associated requirements. Hence, needs transition to requirements through directives and Capabilities. 

In the following section we discuss how the relationship between needs, directives, Capabilities and requirements, as defined by the CE process, aids in pre-RS traceability and conceptually prove that Capabilities help bridge the complexity gap. 

\section{Role of CE in Pre-RS Tracing}
\label{sec:CapabilitiesEngineering}

Traditional RE transitions directly from needs to requirements, as illustrated in Figure \ref{fig:Fig-ComplexityGap}. Needs are the primary source of information for system development. They are stated in a natural language form, and thereby, can often be ambiguous, vague and misleading. Hence, needs are unsuitable as input to the design phase; instead, we utilize system requirements derived from user needs. Thus, there is a transition from needs to requirements. These requirements are more formal, and display quality characteristics such as accuracy, unambiguity, testability, and others \cite{IEEE1998}. Although, formalization of requirements does reduce the possibility of misinterpretations, it usually fails to convey pertinent problem domain information. In fact, it has been recognized that the informality of a natural language has the advantage of communicating certain knowledge that formalization neglects to capture \cite{Goguen1996}. This loss of information has been identified as a key issue in RE \cite{Zave1996}.

We claim that there exists an intermediary space, which symbolizes a middle-ground between the the extremes of formality and informality. This space provides an opportunity to metamorphose from the natural informality of the problem space to the rigid formality of the solution space in a more deliberate and systematic manner. In addition, the consequences of a direct leap from the problem domain to the solution domain --- misinterpreting needs, missing requirements, loss of domain information and so forth --- can be mitigated. We term the intermediary space that includes the complexity gap as the transition space. Again, Figure \ref{fig:Fig-ComplexityGap} graphically illustrates how CE uses the transition space to decrease the leap from problem to solution space. In other words, Capabilities assist in a smoother transition from needs to requirements. 

Sections \ref{sec:ProblemSpace}, \ref{sec:TransitionSpace}, and \ref{sec:SolutionSpace} discuss in detail the problem, transition and solution spaces, respectively. In particular, we first define and describe the \textit{elements of each space}, then discuss what \textit{activity of the CE process} is executed, and finally explain how \textit{traceability is achieved} within that space. Lastly, in Section \ref{sec:ConnectingSpaces} we present a unified perspective on the connectivity between the three spaces, and illustrate how pre-RS traceability is automatically supported by the activities of the CE process.

\subsection{The Problem Space}
\label{sec:ProblemSpace}

\begin{figure}[htbp]
\centering
% trim = left bottom right top
\includegraphics[trim = 25mm 195mm 0mm 20mm, clip, width=11 cm]{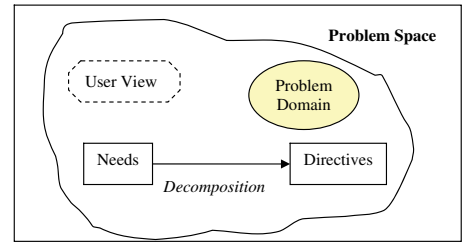}
\caption{\em Problem Space}
\label{fig:Fig-Problem-Space}
\end{figure}

The distinction between a \textit{space} and a \textit{domain} is often blurry and is used interchangeably.
For example, the term problem space is used to indicate a conceptual region of relevance associated with the problem area \cite{Weber2002} \cite{Tracz1995}. However, in some instances the term problem domain is also employed to imply the same \cite{Jackson1995} \cite{Sommerville1997}.  For the purposes of clarity, we utilize the notion of problem and solution domains as discussed by Hull \textit{et al.} to describe spaces \cite{Hull2005}. More specifically, we characterize a space in terms of three elements: the view, the domain and the resident entities. By this characterization, the problem space is composed of the entities: \textit{needs} and \textit{directives}. These entities are defined from a \textit{user's view} and are described in the language of the \textit{problem domain}. Formally, the problem space is a collective aggregation of user view, problem domain, needs, and directives. Activities such as problem identification, decomposition, domain analysis \cite{Diaz1990} and others that help impart a better understanding of the problem, are performed in this space. 
%In addition, the problem space encourages the uninhibited views of the user. 
An application of the user view on the problem domain generates needs, which are, to a large extent, informal and unstructured. A pictorial representation of the problem space is shown in Figure \ref{fig:Fig-Problem-Space}. 
%------------------------------------------------------------------------ 				
% FIGURE: Problem Space
%--------------------------------------------------------------------------
%---------------------------------------------------------------------------- 									
%END FIGURE: Problem Space
%------------------------------------------------------------------------------
\begin{itemize}
	\item \textit{User View:} This refers to the perspective of a stakeholder who is interested in the system to be developed. The user view includes both direct and indirect viewpoints \cite{Kotonya1996}. 
	
	\item \textit{Problem Domain:} The problem domain denotes the knowledge area(s) relevant to the problem being solved. For example, if the problem is to build an ATM, then the problem domain is banking. 

	\item \textit{Needs:} A need specifies what is desired of the system from a user's perspective, and is stated in the language of the problem domain. It is obtained by the applying a user view on the problem domain. It is composed of objects and operations of the problem domain as perceived by the user. Resulting needs cause the generation of other needs. User views can also activate other user views. Hence, every element feeds into the problem space as an input to produce more needs.

	\item \textit{Directives:} We define a directive as a detailed characteristic of the system formulated in the problem domain language. It can be regarded as a requirement with context information. In the problem space, the purpose of a directive is two-fold. First, it helps capture domain information. Second, it facilitates the progress from the problem space to the transition space.	
\end{itemize}

An example of a user need, a directive, a Capability and a requirement, for a hypothetical course management system is described in Table \ref{tab:Differences}.
%------------------------------------------------------------------------ 											
%TABLE: Difference between Elements
%--------------------------------------------------------------------------
\begin{table}[h]
%\centering
\begin{tabular}{|l|l|}
\hline
\textbf{Entity} & \textbf{Example}\\
\hline \hline
Need & \parbox[t][0.25in]{2.2in}{\textit{Need a facility for students and faculty to share ideas, discuss questions}} \\
\hline
Capability & \parbox[t]{2.2in}{\textit{Discussion Forum}} \\
\hline
%Directive & \parbox[t][0.25in]{2.4in}{\textit{Facilitate offline participation in the discussions}}\\
Directive & \parbox[t][0.25in]{2.2in}{\textit{Provide a separate section for faculty to post important announcements}}\\
\hline
A Requirement & \parbox[t][0.4in]{2.2in}{\textit{For the announcement section, the write permission must be enabled only for users designated as faculty}.}\\
\hline
\end{tabular}
\caption{\em Examples in the context of a Course Management System}
\label{tab:Differences}
\end{table}
%------------------------------------------------------------------------ 											
%END TABLE: Difference between Elements
%-------------------------------------------------------------------------

We begin the process of CE by discovering, eliciting and understanding needs from the user. The ``needs'' component of the CE process, and the subsequent \textit{decomposition} to directives illustrated in Figure \ref{fig:FigNeedsToRequirements} is confined to the problem space. In the following section, we explain the decomposition activity, expand on the role of directives and then discuss how the deliverable of the decomposition activity --- a Function Decomposition (FD) graph --- aids traceability.

\subsubsection{CE Activity} 
\label{sec:CE-Decomposition}

\begin{itemize}
	\item \textit{Decomposition:}
Decomposition is an intuitive process of recursively partitioning a problem until an atomic level is reached. This activity uses an FD graph to represent a decomposition of system functionality, as indicated by user needs at various levels of abstraction. We begin with user needs because they help determine what problem is to be solved; in the context of software engineering this means \textit{what} functionality is expected of the system to be developed. Different techniques such as interviews, questionnaires, focus groups, introspection and others \cite{Goguen1993} are employed to gather information from users. A need can be expressed as a real want or a perceived solution. In either case, we are interested only in understanding the functionality expected of the system. Often, because of the informality of the problem domain language, needs are expressed at varying levels of abstraction; the variance in abstraction is depicted by the depth of a node in the FD graph, where depth is the distance of a node from the root. Thus, greater the distance, lower is the abstraction level. The links (edges) between the nodes indicate decomposition, intersection or refinement relationships. The FD graph provides a visual representation of the system functionality, depicting the dependency links between different functions. The use of such a graph supports comprehensive pre-RT traceability as discussed in Section \ref{sec:NeedsTraceability}. 

An FD graph, $G=(V,E)$, is an acyclic directed graph, where $V$ is the vertex set and $E$, the edge set.
Figure \ref{Fig-Example-Graph} presents an example graph. The root node represents the overall mission of the system, internal nodes denote functional abstractions and the leaves symbolize directives. In fact, it has been observed that in a decomposition tree, detailed information is usually specified at the leaves \cite{Weber2002}. Similarly, directives are low-level details pertaining to the system to be developed. An edge can represent decomposition, intersection or refinement of a functionality, as depicted in Figure \ref{Fig-Example-Graph}. Decomposition implies that the functionality of a parent node is equivalent to the union of the functionalities of its children nodes. An intersection edge indicates a common functionality. Refinement is used to elaborate the functionality of the parent node. For a more formal exploration of the FD graph see \cite{Ravichandar2007}.
%------------------------------------------------------------------------ 		
%	FIGURE: Example Graph
%--------------------------------------------------------------------------
\begin{figure}[htbp]
\centering
% trim = left bottom right top
\includegraphics[trim = 10mm 170mm 0mm 15mm, clip, width=9.5 cm]{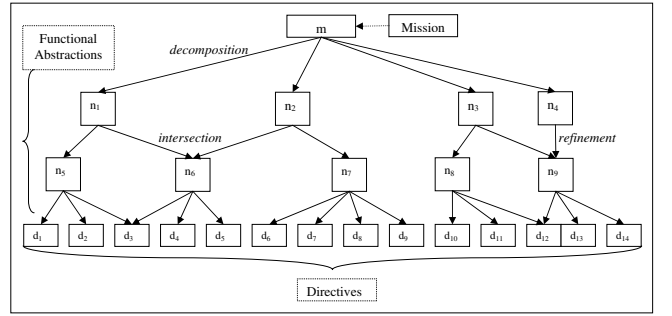}
\caption{\em Example FD Graph $G=(V,E)$}
\label{Fig-Example-Graph}
\end{figure}
%---------------------------------------------------------------------------- 
%END FIGURE: Example Graph
%---------------------------------------------------------------------------
\end{itemize}

\subsubsection{Traceability in the Problem Space}
\label{sec:NeedsTraceability}

We now enumerate how the FD graph, resulting from the activity of decomposition, assists in pre-RS traceability:
\begin{enumerate}
	\item By the virtue of construction, the FD graph represents, in some sense, dependency links between user needs. A change in a parent node affects one or more of its children. Therefore, the effect of a change in a need is more easily traced to its dependent constituents. Subsequently, the FD graph facilitates \textit{traceability among needs}. In contrast, present traceability methods are more concerned with the traces between requirements and its sources, rather than links within the sources themselves. As a result, a changed need is traced only to its associated set of requirements; there is insufficient information to assess the impact of change on other needs, and subsequently, their requirements. Thus, we claim that this type of traceability is important to understand the ripple-effect of change within needs, so as to completely record the effect on requirements. 
When utilizing the FD graph, explicit efforts to maintain traces among the source needs are unnecessary.
	
	\item The FD graph presents an intuitive decomposition of functionality expected of the system to be developed. In the process of decomposition, as shown in Figure \ref{Fig-Example-Graph}, nodes are described at different levels of abstraction. This kind of a visual representation encourages the user to focus on the functionality of the system, rather than on the low-level details. In addition, the hierarchical decomposition of the needs' functionality permits a more systematic approach to stating what is expected of the system. This reduces the possibility of representing conflicting statements, redundant functions, and inconsistent expectations across different user classes. Furthermore, the FD graph, structures the informality of problem space, and thereby, provides an opportunity to automate the traceability process.
	
	\item The leaves of the FD graph are directives, which are described in the language of the problem domain. 
	%They are obtained by the decomposition of user needs. 
	The set of all directives represents the entire system functionality. Figure \ref{fig:FigNeedsToRequirements} indicates that directives are used to formulate Capabilities that occur in the transition space. More specifically, directives are the connecting links between the problem and transition spaces, and thereby, help preserve and transfer the problem domain information from the former to the latter.  We claim that directives implicitly aid pre-RS traceability, by serving as inter-connections between the disparate spaces. Section \ref{sec:TransitionSpace} elaborates on this aspect.
\end{enumerate}

It is inevitable that a large part of the needs elicited from different sources is inconsistent or conflicting \cite{Hunter1997}.However, the use of an FD graph helps capture needs and represent desired system functionalities in a \textit{systematic} manner. Furthermore, the graph also represents functions at different levels of abstraction, which aids in understanding their relative importance. For example, fulfilling the overall mission of the system is more important than implementing a directive. The structure of the graph facilitates traceability among needs by introducing aspects of formality in the highly irregular problem space. 

Recall that the leaves of the graph are directives, which connect the problem and the transition spaces. In the following section, we describe how Capabilities, defined in the transition space, are formulated from these directives, and subsequently, discuss traceability within and between Capabilities.

\subsection{The Transition Space}
\label{sec:TransitionSpace}
\begin{figure}[htbp]
\centering
% trim = left bottom right top
\includegraphics[trim = 25mm 180mm 20mm 20mm, clip, width=10 cm]{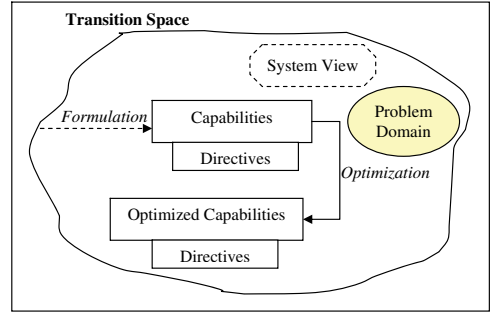}
\caption{\em Transition Space}
\label{fig:Fig-Transition-Space}
\end{figure}

In the transition space, we introduce specific design characteristics of cohesion and coupling. We know that entities with these properties localize and minimize the propagation of change, which is crucial for promoting change-tolerance and also for structuring pre-RS traceability. We move away from the informality of  the problem space by adopting a \textit{system view} --- introducing design aspects of cohesion and coupling --- on the expected functionality of the system. However, we still utilize the \textit{problem domain} but generate new entities termed \textit{Capabilities}. Thus formally, the transition space is defined as a collective aggregation of system view, problem domain, and Capabilities. A pictorial representation of the transition space is shown in Figure \ref{fig:Fig-Transition-Space}. Note that directives are also present in the transition space because they facilitate the change from the problem to the transition space. They originate, however, in the problem space, and so belong there.
%------------------------------------------------------------------------ 				
% FIGURE: Transition Space
%--------------------------------------------------------------------------
%---------------------------------------------------------------------------- 									
%END FIGURE: Transition Space
%------------------------------------------------------------------------------
\begin{itemize}

\item \textit{System View:}  We define system view as the software engineering perspective that guides the identification of Capabilities based on the design principles of high cohesion, low coupling, balanced abstraction levels, as constrained by schedule and technology. Figure \ref{fig:Fig-Transition-Space} illustrates that initial Capabilities are formulated from directives. These initial Capabilities are iteratively optimized to produce an optimized set of Capabilities. In essence, formulation and optimization, the activities of the CE process as shown in Figure \ref{fig:FigNeedsToRequirements}, incorporate the system view to produce Capabilities.
		
\item	\textit{Capabilities:} We describe Capabilities as functional abstractions that exhibit high cohesion, low coupling and are defined at balanced levels of abstraction \cite{Ravichandar2007a}. The optimized set of Capabilities, chosen to develop the system, reflect the constraints of technology feasibility and implementation schedules.
\end{itemize}

The input to the transition space is an FD graph, which describes expected system functionalities at varying levels of abstraction, and directives that help realize these functions. The activity of formulation, uses these directives to identify initial sets of highly cohesive, minimally coupled Capabilities. Then, the optimization activity applies the constraints of schedule and technology to determine the final set of Capabilities. We briefly describe each of these activities and then discuss how they assist in pre-RS traceability. 

\subsubsection{CE Activity} 
\label{sec:CE-FormOptimization}

\begin{itemize}
	\item \textit{Formulation:}
The objective of the formulation activity, as shown in Figure \ref{fig:FigNeedsToRequirements}, is to identify sets of Capabilities from  all possible functional abstractions, depicted in the FD graph. 
For example, for the graph described in Figure \ref{Fig-Example-Graph}, the set of nodes $\{n_1, n_7, n_3\}$ is a potential set of Capabilities because it encompasses all directives. Furthermore, each node is associated with a unique set of directives; directives $\{ d_1,\ldots, d_5\}$ are associated with $n_1$, $\{ d_6,\ldots, d_9\}$ with $n_7$, and $\{ d_{10},\ldots,$ $d_{14}\}$ with $n_3$. Many different sets can be computed from a single FD graph. 
By applying specific measures \cite{Ravichandar2007} we choose those sets which exhibit high cohesion and low coupling values. A discussion about the details of these metrics is outside the scope of this paper. Instead, for the purpose of understanding traceability aspects, we direct our attention to the cohesion and coupling characteristics of Capabilities. 

Cohesion, depicts the ``togetherness'' of elements within an entity. Every element of a highly cohesive unit is directed toward achieving a single objective. Capabilities are designed to possess high functional cohesion, the highest level of cohesion \cite{Bieman1994} among all the other levels 
\cite{Yourdon1979}, and therefore the most desirable. In a Capability all of its associated directives are devoted to realizing the function represented by the Capability. Failure to implement a directive can affect the functionality of the associated Capability with varying degrees of impact. We hypothesize that the degree of impact is directly proportional to the relevance of the directive to the functionality. Consequently, the greater the impact, the more essential the directive. Hence, each directive is assigned a weight on a $[0,1]$ scale to indicate its relevance value; existing risk impact categories: Catastrophic, Critical, Marginal and Negligible \cite{Boehm1989} are used to guide this assignment \cite{Ravichandar2007}.

The other characteristic of Capabilities is low coupling. Coupling is a measure of interdependence between entities \cite{Stevens1974}. Dependency links between entities behave as change propagation paths. The higher the number of links, the greater is the likelihood of ripple effect.  
The edges of the FD graph indicate the links between Capabilities and their directives. We measure coupling between Capabilities as a function of the coupling between their constituent directives \cite{Ravichandar2007}.

\item \textit{Optimization:}
The inputs for the optimization activity are sets of highly cohesive, minimally coupled Capabilities.
This activity aims to identify that set which best accommodates the constraints of schedule and technology, and is an ongoing component of our current research. 
\end{itemize}

\subsubsection{Traceability in the Transition Space}
\label{sec:TraceabilityTransition}
We now discuss how the structure and characteristics of a Capability and its associated directives assist in pre-RS traceability in the transition space. 
\begin{enumerate}
\item Capabilities are functional abstractions identified from the FD graph, and are associated with a set of directives. Capabilities provide stakeholders with a high-level perspective of the system functionality. In contrast, directives within each Capability furnish the more rudimentary details. We claim that the structure of Capabilities and their directives serve the pre-RS traceability interests of two disparate groups of users --- high-end and low-end users --- as identified by Ramesh \cite{Ramesh1998}. High-end users work with complex systems that have a large number of requirements, and can easily become entangled in the details. However, Capabilities can be utilized to understand, from a high-level perspective, the expected system functionalities, and thereby, ensure that user expectations are satisfied. In contrast, low-end users are known to neglect pre-RS traceability because they are more concerned with detailed requirements. In such a case, these users can focus on directives to trace the origin of requirements because they are similar to detailed requirements but are stated in the language of the problem domain. Thus, we claim that Capabilities and their directives help overcome certain shortcomings of pre-RS traceability among different groups of users.

\item Capabilities exhibit high functional cohesion. Each associated directive, with varying degrees of relevance, is essential for realizing a Capability. A change in a directive may affect the parent Capability, and also other associated directives. Thus, Capability-directive links established by the FD graph can be utilized for the purposes of traceability. In other words, the property of high cohesion facilitates traceability within a Capability. 

\item Minimally coupled Capabilities reduce the overhead of maintaining traceability information between each and every directive. This is because low coupling implies decreased dependencies, and therefore, reduces the need for traceability paths between certain entities.

\item One can use information from the FD graph to record which directives are common to different Capabilities, and maintain traces for them. This is in agreement with the observation that a change in a detailed requirement is often less significant, and hence, it is more cost efficient to maintain traceability for critical entities \cite{Ramesh1998}. The criticality of directives can be deduced from their relevance values, which are elicited for the computation of the cohesion measure.
\end{enumerate}
In the transition space, we use the system view to formulate and optimize Capabilities from the FD graph generated in the problem space. We observe that directives behave as linkages between the two spaces, and  also, preserve and transfer problem domain knowledge. These directives are transformed to requirements, which belong to the solution space. We now discuss the solution space, the CE activity of transformation (shown in Figure \ref{fig:FigNeedsToRequirements}) and traceability within the space.

\subsection{The Solution Space}
\label{sec:SolutionSpace}
\begin{figure}[htbp]
\centering
% trim = left bottom right top
\includegraphics[trim = 10mm 190mm 40mm 20mm, clip, width=10 cm]{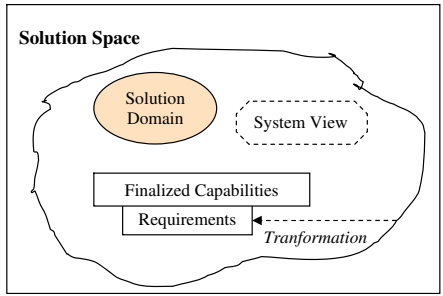}
\caption{\em Solution Space}
\label{fig:Fig-Solution-Space}
\end{figure}

The solution space is usually understood as the conceptual realm associated with the technical aspects of developing the system. Often, as with the problem space, the terms solution domain and space are used interchangeably. However, we envision the solution space as being defined by the \textit{system view}, which generates entities called \textit{requirements} from directives. These requirements are described in the language of the \textit{solution domain}. We formally describe the solution space as the collective aggregation of the system view, the solution domain, and requirements. In the solution space, activities related to developing the system, such as establishing requirements, modeling architecture, developing design specifications, coding, unit testing, integration and testing, system maintenance and other downstream development processes are performed. We restrict our focus up to the specification of requirements, and therefore, the graphical representation of the solution space in Figure \ref{fig:Fig-Solution-Space} presents only Capabilities and requirements as its entities. Capabilities that are present in the solution space, are the unchanged entities that originated in the transition space. Hence, Capabilities behave as connectors between the transition and solution spaces, just as directives do between the problem and transition spaces.
%------------------------------------------------------------------------ 				
% FIGURE: Solution Space
%--------------------------------------------------------------------------
%---------------------------------------------------------------------------- 									
%END FIGURE: Solution Space
%------------------------------------------------------------------------------

The element, system view, has already been defined in Section \ref{sec:TransitionSpace}. Therefore, we now describe the other elements --- solution domain and requirements --- of the solution space.
\begin{itemize}
	\item \textit{Solution Domain}: Solution domain denotes the technical area(s) relevant to the system being developed. This assumes that a solution always implies the development of a system, which is true in  software engineering. For example the solution domain provides technical concepts such as design patterns and architectural styles that are relevant to the design of an ATM.	

\item \textit{Requirements}:  A requirement is a statement specified in the language of the solution domain that states what is expected of the system and exhibits specific quality characteristics such as testability, verifiability, accuracy, unambiguity, and others \cite{IEEE1998}. 
\end{itemize}

The input to the solution space is an optimized set of Capabilities and their associated directives. Moving from the transition to the solution space entails a change in the domain language used to describe the entities. In particular, requirements are now stated in the solution domain language. These requirements are derived from directives. Recall, directives originate in the problem space, assist the change to the transition space and are now transformed to requirements in the solution space. 
Also, Capabilities identified in the transition space progress unchanged into the solution space, and in the process become associated with a set of requirements as opposed to a set of directives. We now examine the transformation activity and discuss traceability in the solution space.
	
\subsubsection{CE Activity}
\begin{itemize}
\item \textit{Transformation:} 
Requirements are still the basis for developing systems. They fulfill several purposes that include providing the rationale for design, criteria for validation and others as enumerated in \cite{Parnas1986}. Thus, there is a need to transform directives to requirements.
In systems that are incrementally developed, only the directives associated with the Capability chosen for development are to be transformed to requirements. This is in agreement with the principle of delaying the specification of requirements in Capability Based Acquisition \cite{Montroll2003}, and thereby, avoiding the pitfalls of fixed requirements. We hypothesize that there is a one-many mapping between directives and requirements.
\end{itemize}

\subsubsection{Traceability in the Solution Space}
\label{sec:TraceabilityInSolutionSpace}
Traceability aspects of the transition space are applicable to the solution space because of the similarity between directives and requirements. Both are low-level detailed entities that describe what is expected of the system. We enumerate the traceability advantages in the solution space obtained by the utilization of the CE process.
\begin{enumerate}
\item When transforming directives to requirements, one can use the relevance values to determine the criticality of requirements. Recall that relevance values indicate the importance of a directive in achieving the objective of its parent node. For example, a directive with a relevance value $1$ is mission-critical, and therefore, requirements derived from this directive are most likely critical too. This information can be utilized to selectively capture certain traces. In fact, it has been recognized that not all requirements are equal and that it is cost-effective to maintain traces to and from critical requirements \cite{Ramesh1998}.
	
\item Pinheiro \cite{Pinheiro2003} describes \textit{inter-requirements traceability} as capturing the relationships between requirements. The CE process assists this type of traceability in three different ways:

\begin{enumerate}
	\item A directive is transformed into one or more requirements. Because the requirements are derived from the same directive, they share a very strong relationship, and therefore, a change in one is most likely to affect the other requirements. Hence, there is an implicit inter-requirements traceability among requirements derived from the same directive source. 

	\item In the solution space, Capabilities are associated with a set of requirements that are transformed from directives. Note that Capabilities are unchanged when progressing from the transition to the solution space. By definition, Capabilities exhibit high functional cohesion; every element is essential to attaining its objective. Therefore, requirements associated with each Capability are strongly related to each other because  each requirement is working towards fulfilling the same Capability. This facilitates inter-requirements traceability  within a Capability and alleviates the overhead of analyzing exponential number of relationships among all possible requirements.

\item Minimally coupled Capabilities aid in selective traceability between the requirements associated with different Capabilities. Directives that are shared among Capabilities in the transition space result in requirements which are common in the solution space. As a result, traceability efforts can focus on these requirements, which have the potential to affect more than one Capability when changed. 

\end{enumerate}
	
\item Tracing is performed only when a need is perceived. For example, requirements are tagged with keywords, cross-referenced, etc., to facilitate future tracing. However, the same importance has not been extended to the tracing from needs to requirements. We conjecture from the observations above that the process of CE may ease the difficulty of tracing from needs to requirements and thereby, further assist pre-RS traceability.
\end{enumerate}

Thus, in the solution space, requirements are specified for Capabilities that are to be developed. These requirements are obtained from directives by the activity of transformation. The nature of a Capability, and its directives or requirements, facilitate traceability in the transition space and the solution space, respectively. Moreover, the properties of Capabilities assist in inter-requirements traceability. We now briefly discuss, from a more global perspective, how the different spaces are connected.

\subsection{Connecting Spaces}
\label{sec:ConnectingSpaces}

\begin{figure}[h]
\centering
% trim = left bottom right top
\includegraphics[trim = 5mm 130mm 0mm 20mm, clip, width=9 cm]{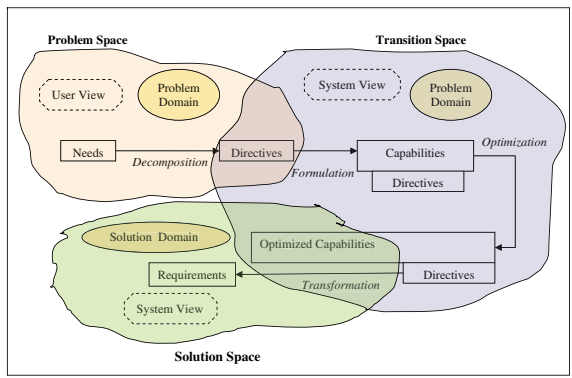}
\caption{\em Capabilities Engineering in terms of Spaces}
\label{fig:Fig-All-Spaces}
\end{figure}

The preceding sections have described and discussed the traceability aspects in each space. In this section, we adopt a unified perspective and examine the relationship between the spaces. This is essential to understand the potential for traceability from needs to requirements using the CE process. When making a transition from one space to another, either the domain or the view varies.
%We term the element that varies as a change agent. i.e. exactly one change agent is responsible for the transition. 
For example, the \textit{view} changes in the progression from the problem space to the transition space;  the \textit{domain} changes in the shift from transition space to the solution space. This is presented in Table \ref{tab:SolutionSpace}, where the italicized entries indicate the changed element.
%------------------------------------------------------------------------ 											
%TABLE: All Spaces
%--------------------------------------------------------------------------
\begin{table}[h]
\centering
\begin{tabular}{|c | c |c | c |}
\hline
\textbf{Space} & \textbf{Domain} & \textbf{View} & \textbf{Entities} \\
\hline \hline
\textbf{Problem} & Problem  & User & Needs, Directives \\
\hline
\textbf{Transition} & Problem  & \textit{System} & Capabilities\\
\hline
\textbf{Solution} & \textit{Solution} & System  & Requirements \\
\hline
\end{tabular}
\caption{\em Connecting the Spaces}
\label{tab:SolutionSpace}
\end{table}
%------------------------------------------------------------------------ 											
%END TABLE: All Spaces
%-------------------------------------------------------------------------
In addition, we observe that the spaces are connected through common entities. Needs are decomposed to directives in the problem space; these directives are utilized to identify Capabilities in the transition space. As mentioned earlier, directives carry with them problem domain information which is 
preserved in the transition space. Furthermore, they are also used to identify Capabilities, which pass unchanged into the solution space. Thus, Capabilities and directives behave as connectors between the problem and solution spaces. Figure \ref{fig:Fig-All-Spaces} graphically illustrates the transitions between and common elements resident in the spaces.
%------------------------------------------------------------------------ 				
% FIGURE: All Spaces
%--------------------------------------------------------------------------
%---------------------------------------------------------------------------- 									
%END FIGURE: All Spaces
%------------------------------------------------------------------------------
The solution approach of the CE process illustrated in Figure \ref{fig:Fig-All-Spaces} strives to resolve the problem of the complexity gap represented in Figure \ref{fig:Fig-ComplexityGap}. This is achieved by  introducing Capabilities to bridge the chasm between the spaces. In addition, the conventional overhead and difficulties associated with pre-RS traceability are alleviated with the utilization of the CE process.

\section{Traceability Challenges}
\label{sec:AddressingSpecificGCTChallenges}

In this section we examine from a pre-RS traceability perspective how CE addresses specific challenges. In particular, we refer to the challenges enumerated in the Grand Challenges document \cite{Traceability2006} to structure our discussion.

\subsection{Supporting Evolution}
\label{sec:SupportingEvolution}
Large-scale systems constantly evolve during their lengthy development cycles. Consequently, there is an enormous overhead in ensuring that the traceability links depict current dependencies.  Changes can occur in needs, directives, Capabilities or requirements. However, we are specifically concerned with the impact of any such change on Capabilities or requirements because they are the formal inputs to the development phases. Therefore, we present how CE facilitates traceability to requirements in the event of different change scenarios.

\begin{itemize}
	\item \textit{Needs Change:} The FD graph illustrates a decomposition of functions, which essentially represents the user needs. The visual structure of the graph provides information about nodes that can be affected by a need change. For example, in Figure \ref{Fig-Example-Graph} if a need associated with node $n_3$ changes, then it is evident that nodes $n_8$ and $n_9$ are most likely to be affected because they are connected by decomposition edges to the parent node $n_3$. If any of the affected nodes are Capabilities, then the set of associated requirements are also impacted. 
	 	
	\item \textit{Directives Change:} In the event of a directive change, its relevance value can provide an estimate of the possible impact on the associated Capability. Furthermore, this change also affects those requirement(s) obtained from the changed directive.
Also, coupling between Capabilities is computed as a function of the coupling between their respective directives. Therefore, low coupling implies that a change in a directive has a decreased likelihood of affecting directives in another Capability, or their derived requirements.
		
	\item \textit{Capabilities Change:} The high cohesion and low coupling of a Capability contains the impact of change to within a Capability. In all likelihood, only the set of requirements associated with it are affected. The dependencies between Capabilities can be examined using the FD graph to analyze the paths of change propagation.
	
	\item \textit{Requirements Change:} As with directives, requirement changes are contained to within a Capability. Reduced coupling between Capabilities ensures that a change in a requirement has low impact on requirements associated with other Capabilities. 
\end{itemize}
	
Thus, no special effort is required to maintain the traceability information about all links in the pre-RS phase. Instead, the design of the FD graph and the subsequent derivations of requirements from directives, provide a natural means for traceability. In addition, a Capability's characteristics of high cohesion and low coupling, contain and reduce the impact of change, respectively. This serves to alleviate the burden of capturing all possible traces, which is otherwise difficult in an evolving system.  

\subsection{Link Semantics}
\label{sec:LinkSemantics}
The FD graph is constructed in the problem space in accordance to certain rules. In particular, different types of nodes and edges are interpreted in the context of the problem space. As a result, by the virtue of its definition, the FD graph provides both link semantics and indicates the granularity of nodes. This information is essential to understand the underlying traceability relationships.
\begin{itemize}
	\item \textit{Link Type:} The FD graph is currently concerned only with representing expected system functionality, and not with the depiction of non-functional needs. Therefore, we define the edges of the graph to represent decomposition, intersection or refinement relationships. These links are independent of any particular domain, and therefore, are fundamental in nature. 
	
	\item \textit{Granularity:} The root of the graph represents the overall system mission. In contrast, the leaves denote the directives, which are low-level detailed characteristics. The internal nodes occupy the middle ground depicting functionalities at different levels of abstraction. We observe that the abstraction level of a node increases with its decreasing distance from the root (distance is the number of edges in the shortest path). Thus, the natural hierarchical structure of the graph indicates the granularity of different entities.
	
\end{itemize}

\subsection{Scalability}
\label{sec:Scalability}

Current traceability techniques are more suited for tracing from and among structured documents of the solution space, than within the informality of the problem space. In addition, these techniques also have to scale up to maintaining traceability in large systems. Our solution approach, the CE process, 
% utilizes basic principles of software engineering to design change-tolerant systems. Its 
provides inherent traceability information, which can be used to adequately manage the enormous complexity of large-scale systems, or to serve the interests of small-scale system development. In either case, the CE approach facilitates traceability within and between the different spaces. In particular, as discussed in Section \ref{sec:NeedsTraceability}, CE facilitates traceability in the problem space, which has been largely neglected hitherto.

\subsection{Human Factors}
\label{sec:HumanFactors}

Traceability is often viewed as an activity that is extraneous to the actual development process. Human beings regard traceability efforts as invasive and time-consuming because of the inability to generate automatic traces as a by-product of development. In addition, it is difficult to trace between artifacts created by different users. However, we claim that using a Capabilities-based approach reconciles these problems:
\begin{itemize}
	\item \textit{Integrating Traceability:} The FD graph is the basis for the decomposition and formulation activities of the CE approach. Additionally, its edges behave as links, and thereby serve as a means for tracing between needs, directives and Capabilities. Thus, unlike current techniques in the CE development approach there is little overhead in producing trace information.
	
	\item \textit{Bridging Semantic Differences:} As discussed in Section \ref{sec:ConnectingSpaces}, the difficulty of tracing artifacts across different spaces is alleviated by maintaining the linkages between spaces.
\end{itemize}
 
\subsection{Cost Benefit Analysis}
\label{sec:CostBenefitAnalysis}
Complete and comprehensive traceability between every entity produced during the development process is theoretically desirable but practically infeasible. The cost of maintaining all possible traces is not commensurate with the advantages one may obtain. In particular, as discussed in Section \ref{sec:TraceabilityInSolutionSpace}, to derive maximal benefits of traceability it is more prudent to maintain links between entities that are mission-critical or those that are described at a high-level of abstraction. We know that Capabilities enable one to abstract back from the lower details and focus on larger aspects of complex systems. By the same token, the  requirements associated with Capabilities permit the analysis of details pertinent to that function. 

We certainly acknowledge that there is an overhead cost associated with CE. We conjecture that this cost is minimal when one considers that the inherent traceability provided by CE reduces the cost associated with upfront traceability effort. Additionally, the cost-savings achieved through the property of change-tolerance, the ease of downstream maintainability, and the facility to support critical traceability, argue that the the introduction of Capabilities provides benefits that exceed the costs of traceability.

\section{Conclusion}
\label{sec:Conclusion}

Empirical evidence suggests that pre-RS traceability is crucial for the success of RT. This type of traceability is concerned with capturing relationships between requirements and their sources, which are primarily user needs. However, this process is challenged by the vast disparity between the informality of a user need and the rigidity of a system requirement. 
We claim there exists an intermediary space called the transition space, which can structure the movement from one space to the other. More specifically, we identify highly cohesive, minimally coupled, optimized functional abstractions termed Capabilities in this space. Here the Capabilities are associated with a set of directives, which are obtained by the decomposition of user needs. Hence, directives preserve and carry domain information from the problem to the transition space. However, it is the Capabilities that progress unchanged into the solution space, where their directives are transformed to requirements. Therefore, although, the spaces are dissimilar they are connected through common entities. By establishing such relationships, we are no longer constrained by the traditional techniques of traceability. The use of directives and Capabilities generates an inherent traceability mechanism from needs to requirements. Furthermore, the structured nature of this approach supports the development of automated tools to capture and analyze different trace paths.
Thus, by the virtue of using a Capabilities-based development approach, the effort to directly relate informal needs and formal system requirements is reduced, the complexity gap between the spaces is bridged, and the ability to trace requirements back to their needs and vice-versa, \textit{i.e.} pre-RS traceability, is provided. \\ 

\bibliography{GCT07}
\end{document}